\documentclass[sigconf]{acmart}

\usepackage{enumitem}

\AtBeginDocument{%
  \providecommand\BibTeX{{%
    \normalfont B\kern-0.5em{\scshape i\kern-0.25em b}\kern-0.8em\TeX}}}

\copyrightyear{2024}
\acmYear{2024}
\setcopyright{acmlicensed}\acmConference[WWW '24 Companion]{Companion Proceedings of the ACM Web Conference 2024}{May 13--17, 2024}{Singapore, Singapore}
\acmBooktitle{Companion Proceedings of the ACM Web Conference 2024 (WWW '24 Companion), May 13--17, 2024, Singapore, Singapore}
\acmDOI{10.1145/3589335.3651242}
\acmISBN{979-8-4007-0172-6/24/05}

\begin{document}

\title{RecAI: Leveraging Large Language Models for Next-Generation Recommender Systems}

\author{Jianxun Lian}
\email{jialia@microsoft.com}
\affiliation{%
  \institution{Microsoft Research Asia} 
  \city{Beijing}
  \country{China}
}

\author{Yuxuan Lei}
\email{leiyuxuan@mail.ustc.edu.cn}
\affiliation{%
  \institution{University of Science and Technology of China} 
  \city{Hefei}
  \country{China}
  \postcode{230031}
}

\author{Xu Huang}
\email{xuhuangcs@mail.ustc.edu.cn}
\affiliation{%
  \institution{University of Science and Technology of China} 
  \city{Hefei}
  \country{China}
  \postcode{230031}
}

\author{Jing Yao} 
\email{jingyao@microsoft.com}
\affiliation{%
  \institution{Microsoft Research Asia} 
  \city{Beijing}
  \country{China}
}

\author{Wei Xu} 
\email{xu_wei@ruc.edu.cn}
\affiliation{%
  \institution{Renmin University} 
  \city{Beijing}
  \country{China}
}


\author{Xing Xie}
\email{xingx@microsoft.com}
\affiliation{%
  \institution{Microsoft Research Asia} 
  \city{Beijing}
  \country{China}
}

\renewcommand{\shortauthors}{Jianxun and Yuxuan, et al.}

\begin{abstract}
This paper introduces RecAI, a practical toolkit designed to augment or even revolutionize recommender systems with the advanced capabilities of Large Language Models (LLMs). RecAI provides a suite of tools, including Recommender AI Agent, Recommendation-oriented Language Models, Knowledge Plugin, RecExplainer, and Evaluator, to facilitate the integration of LLMs into recommender systems from multifaceted perspectives. 
The new generation of recommender systems, empowered by LLMs, are expected to be more versatile, explainable, conversational, and controllable, paving the way for more intelligent and user-centric recommendation experiences. We hope the open-source of RecAI can help accelerate evolution of new advanced recommender systems. The source code of RecAI is available at \url{https://github.com/microsoft/RecAI}.
\end{abstract}



\maketitle

\section{Introduction} 
Large language models (LLMs) have been rigorously pretrained on massive amounts of data sourced from the internet. With the expansion of their model parameters from the hundreds of millions to the hundreds of billions, LLMs have demonstrated emerging general intelligence, such as engaging in smooth conversations, executing logical and mathematical reasoning, following detailed instructions to complete tasks, and assisting in the troubleshooting of software development issues. Consequently, a diverse set of applications is now transitioning toward the integration of LLMs, either to bolster existing models or to implement them as the principal framework.

Recommender systems (RSs) function as a specialized type of information retrieval system, designed to capture a user's preferences from their profile and behavioral history. RSs can curate a selection of items to present to the user, thereby simplifying the process of discovering preferred choices within an extensive database of items. Impressed by the remarkable ability of LLMs, there is burgeoning interest in how LLMs can transform the landscape of next-generation RSs. However, directly applying LLMs as recommender models is not feasible. On one hand, the knowledge boundary of LLMs is limited to the information available up to the point of their last training update.  The specific item catalog and the attributes of items within a particular recommendation context may not be fully captured by LLMs. On the other hand, user preference patterns are not only domain-specific but also subject to rapid evolution. Consequently, traditional recommender models require frequent retraining or fine-tuning with up-to-date data to capture the unique and shifting patterns that diverge from the general world knowledge encoded in LLMs.

This paper investigates the possibilities of utilizing LLMs to advance RSs. The vision is for the next wave of recommender systems, empowered by LLMs, to exhibit heightened intelligence and versatility.  This includes the ability to generate explanations for recommendations, facilitate item suggestions through conversational interfaces, and offer enhanced user control. To achieve these objectives, we introduce RecAI, a lightweight toolkit to integrate LLMs into RSs from a comprehensive and diverse set of perspectives. Currently, RecAI comprises five foundational pillars, each one corresponds to an independent application scenario:
\begin{itemize}[leftmargin=*]
    \item \textbf{Recommender AI Agent}. This is an LLM-driven AI agent, where the LLMs act as the "brain" responsible for user interaction, as well as for reasoning, planning, and task execution. Traditional recommender models act as "tools", enhancing the LLMs by providing specialized capabilities.
    \item \textbf{Recommendation-oriented LM}. Fine-tuning language models is an effective strategy for integrating domain-specific knowledge into models. We introduce two types of models: \textit{RecLM-emb} and \textit{RecLM-gen}. RecLM-emb converts diverse text types, such as natural conversations and unstructured attributes, into embeddings for item retrieval. RecLM-gen is a generative language model. After fine-tuned with in-domain data, it excels at understanding domain information and collaborative patterns.
    \item \textbf{Knowledge Plugin}. This supplements LLMs by dynamically incorporating domain-specific knowledge into prompts without altering the LLMs themselves. This is particularly beneficial when LLMs cannot be fine-tuned --- either due to only API availability or constraints like lack of GPU resources. 
    \item \textbf{RecExplainer}. Most deep learning-based recommender models are opaque, acting as "black boxes." RecExplainer is designed to leverage LLMs' ability to elucidate the workings of embedding-based recommender models by interpreting the underlying hidden representations.
    \item \textbf{Evaluator}. RecAI includes a tool for assessing LLM-augmented recommender systems in an convenient manner. It encompasses the evaluation of embedding-based and generative recommendations, explanation capabilities, and conversation abilities.
\end{itemize}
In the following sections, we will introduce details for each pillar.

\section{Recommender AI Agent}\label{sec:agent}
The remarkable achievements of LLMs have inspired researchers to envision a future where RSs are more versatile, interactive, and user-centric. However, the use of LLMs as independent recommender models is constrained by their lack of domain-specific knowledge. Traditional recommender models are tailored to specific recommendation tasks through training on domain-specific data, presenting an opportunity for synergy. Combining the strengths of both LLMs and specialized recommender models into a unified framework emerges as a promising approach. This synthesis is an LLM-based agent framework, wherein recommender models serve as specialized tools for tasks like item retrieval or click-through rate (CTR) prediction, while LLMs operate as the core intelligence, facilitating smooth interactions with users and employing contextual reasoning to determine the most suitable tools for the current conversational context. We name this AI agent framework \textbf{InteRecAgent}~\cite{huang2023recommender}.

We define a core suite of three distinct tool types within InteRecAgent to enable effective communication with users:
(1) \textsl{Information Query}: The InteRecAgent addresses user queries alongside recommending items. For instance, on a gaming platform, it can answer questions about game details like release dates and prices by querying a backend database with SQL.
(2) \textsl{Item Retrieval}: This tool suggests a list of potential items based on a user's criteria. InteRecAgent differentiates between "hard conditions" (explicit user specifications) and "soft conditions" (preferences requiring semantic matching). SQL tools and item-to-item matching based on embeddings are used to fulfill these conditions, respectively.
(3) \textsl{Item Ranking}: Ranking tools predict user preferences on the shortlisted items using user profiles and/or user history, ensuring recommendations align with both the user's immediate needs and their overall preferences. These shortlisted items may either be derived from the item retrieval process or be provided by the user.

\begin{figure}[h]
    \centering
    \includegraphics[width = 0.48\textwidth]{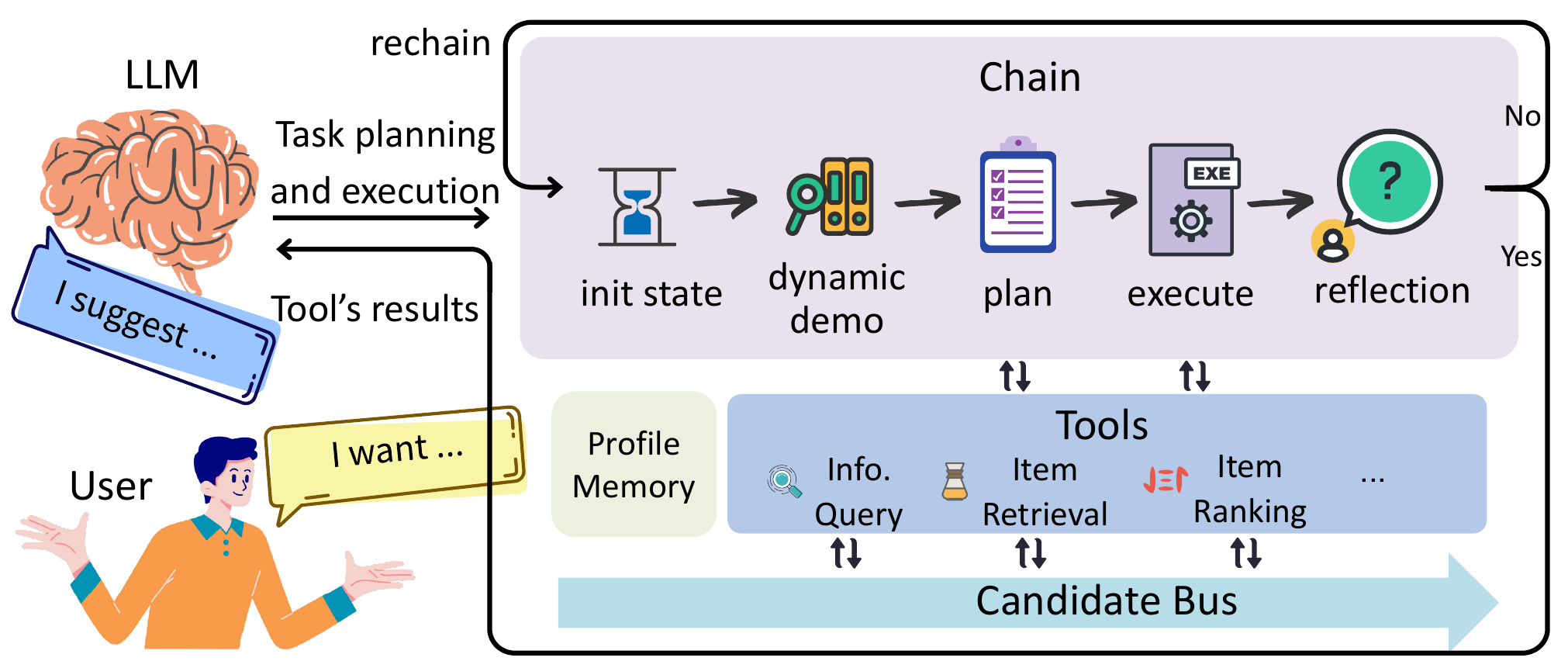}
    \caption{An overview of the InteRecAgent. Users interact with an LLM in natural language. The LLM comprehends users' intention and makes a tool-execution plan to fetch the correct items or information from the specific domain. Based on tools' results, the LLM generate response for users.}
    \label{fig:agent}
\end{figure}

Memory, task planning, and tool-learning are three critical components for AI agents. In InteRecAgent, we also tail these three components to address specific challenges in the recommendation scenario. A simple illustration of InteRecAgent can refer to Figure~\ref{fig:agent}.

\textbf{Memory}. 
To effectively manage the flow of item candidates within InteRecAgent and address input context length limitations, we introduce two key modules: the Candidate Bus and User Profile. The Candidate Bus serves as a dedicated memory system for storing current candidates and tracking tool outputs, facilitating the streamlined processing of item lists and tool execution records. This ensures efficient interaction among tools without burdening the LLM's input prompt. User Profiles are constructed from conversation histories and differentiated into long-term and short-term memories. This segmentation tackles the complexities of lifelong learning scenarios while emphasizing users' immediate requests, allowing for refined and adaptive recommendations.

\textbf{Task Planning}.
We adopt a plan-first approach for the InteRecAgent, diverging from the traditional step-by-step method. Initially, the LLM devises a comprehensive execution plan based on the user's intentions from the dialogue. Subsequently, it strictly follows this plan, sequentially invoking tools that interface through the Candidate Bus. The plan phase incorporates user input, context, tool descriptions, and demonstration for in-context learning to create a tool utilization plan. The execution phase then follows the plan, with each tool's output tracked except for the final output, which informs the LLM's response. To enhance planning, we use dynamic, high-quality demonstrations, selecting examples most similar to the current user's intent. The plan-first approach reduces API calls and latency, crucial for conversational interaction, and improves planning capability with efficient demonstration strategies.

\textbf{Tool-learning}.
In our quest to make the InteRecAgent framework more accessible and cost-effective, we explore the potential of training smaller language models (SLMs) like the 7B-parameter Llama to emulate GPT-4's adeptness at following instructions. We create RecLlama, a fine-tuned version of Llama-7B, using a specialized dataset generated by GPT-4 that contains pairs of [instructions, tool execution plans]. To ensure dataset quality and diversity, we combine data from user simulator-agent dialogues with crafted dialogues covering various tool execution scenarios. We find that RecLlama can significantly outperform some LLMs such as GPT-3.5-turbo and Text-davinci-003 in serving as the brain in InteRecAgent.

Detailed evaluations of InteRecAgent are presented in \cite{huang2023recommender}.

\section{Recommendation-oriented Language Model}\label{sec:finetune}
\begin{figure}[h]
    \centering
    \includegraphics[width = 0.48\textwidth]{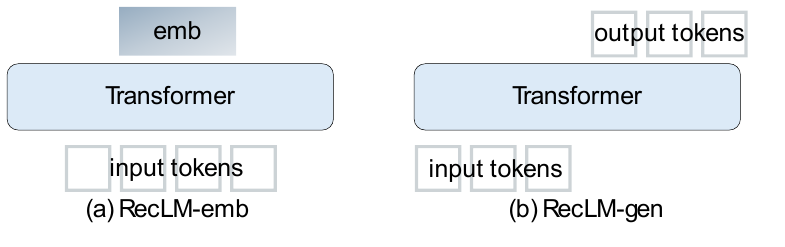}
    \vspace{-4ex}
    \caption{A graphical comparison of two RecLM structures.}
    \label{fig:reclm}
\end{figure}

Traditional RSs typically handle structured data, such as sequences of item IDs, to infer user preferences. However, this structured approach is not well-suited to the strengths of large language models (LLMs), which are adept at processing natural language. In real-world interactions, users often provide a wealth of information in their conversations, ranging from explicit requests to subtle indications of their preferences, articulated in their natural language. LLMs are capable of interpreting these user intents and translating them into natural language-based directives for subsequent processing. Therefore, there's a critical demand for RSs capable of assimilating diverse textual inputs --- from casual dialogues to unstructured product descriptions --- and returning items that closely match the intricacies of the query. To this end, we propose fine-tuning language models specifically for recommendation tasks. Depending on whether the approach is embedding-based or generative, we introduce two distinct types of models: RecLM-emb and RecLM-gen, as illustrated in Figure~\ref{fig:reclm}.

\subsection{RecLM-emb}\label{sec:reclmemb}
Previous research has built general-purpose text embedding models using contrastive pre-training on expansive datasets to enhance semantic text matching. Yet, these models often do not perform well in zero-shot item retrieval tasks. The main issue is their generalized representations, which don't adequately capture the specific details of items mentioned in variously structured queries.

To overcome this limitation, we have crafted ten matching tasks that address different facets of item representation and compiled a fine-tuning dataset tailored for item retrieval. Utilizing this dataset, we introduce our embedding-based Recommendation Language Model~\cite{lei2024aligning}, {RecLM-emb}, designed to retrieve items based on textual input of any form. After fine-tuning, RecLM-emb demonstrates a notable enhancement in performance on item retrieval tasks. It also shows effectiveness in conversational scenarios, thereby enhancing the capabilities of LLM-based recommender agents like Chat-Rec~\cite{gao2023chat}. Moreover, RecLM-emb has the potential to unifying search and recommendation service or for generating refined semantic representations to support downstream rankers.

\subsection{RecLM-gen}\label{sec:reclmgen}
In contrast to embedding-based LMs, the generative recommendation LM, abbreviated as RecLM-gen~\cite{lu2024aligning}, decodes responses directly into natural language. When it comes to recommending items, the names of these items are seamlessly integrated into the dialogue. As such, RecLM-gen manages user-system interactions in an end-to-end fashion, eliminating the need for intermediary steps like embedding-based retrieval or tool invocation.

\cite{hou2023large} reveals that with carefully crafted prompt engineering and bootstrapping techniques, zero-shot LLMs can serve as competent ranking models. Nonetheless, our observations suggest that fine-tuning with domain-specific data can lead to even more substantial improvements in recommendation performance. A fine-tuned 7B Llama-2-chat model can surpass GPT-4 in item ranking tasks. In RecAI, we offer the fine-tuning scripts for RecLM-gen, enabling users to replicate and build upon our results.

The advantages of RecLM-gen are three-fold. Firstly, domain-specific fine-tuning equips the LM to better recognize item names and unique collaborative patterns, thereby surpassing the accuracy of general-purpose LMs in recommendations. Secondly, integrating RecLM-gen as the core intelligence of the Recommender AI Agent framework significantly lowers system costs compared to larger, more costly LMs. Lastly, RecLM-gen facilitates seamless, real-time user interactions by generating tokens streamingly, unlike traditional AI agent frameworks that rely on multiple backend LLM calls for context reasoning and tool interaction, which can introduce delays of 10-20 seconds as per our observations.

\section{Knowledge Plugin}\label{sec:knowledge}
In scenarios where fine-tuning LLMs is not feasible --- due to only having access to LLM APIs or facing constraints in terms of GPU resources or time --- we must find alternative ways to introduce domain-specific knowledge. Notably, the input context window size of LLMs is expanding, as evidenced by GPT-4-turbo's increase to 128k tokens and Claude 2.1's support for up to 200k tokens. This expansion provides an opportunity to include selected domain patterns directly into the input.

Motivated by this, we propose the Domain-specific Knowledge Enhancement (DOKE) paradigm, which bypasses the need for parameter modification and instead uses prompts to integrate domain knowledge. The core idea of DOKE includes three steps: (1) extracting domain relevant knowledge, (2) selecting knowledge pertinent to the current sample to fit within prompt length constraints, and (3) formulating this knowledge into natural language.  

As a instantiation of applying the DOKE paradigm to RSs, we focus on boosting LLMs' performance on the item ranking. Our specialized knowledge extractor gathers item attributes and collaborative filtering signals, tailoring this information to the user's preferences and the set of candidate items. It then conveys this information either through natural language explanations or as reasoning paths on a knowledge graph, thereby yielding more interpretable recommendations.
Through is way, 
our experimental results across different recommendation benchmarks demonstrate that DOKE markedly enhances LLM performance, proving its efficiency and adaptability. For additional details, please refer to \cite{yao2023knowledge}.

\section{RecExplainer}\label{sec:recexplainer}
Model interpretability is crucial for creating reliable RSs, as it provides insights into system reliability, aids in detecting bugs, helps identify biases, and drives innovation. One major approach in this research field is training self-explainable surrogate models to mimic the behavior of more complex models. However, surrogate models tend to compromise model accuracy and typically generate explanations in fixed, less intuitive formats like lists of feature importance or decision rules.

LLMs offer a new perspective for surrogate modeling that avoids a hard trade-off between model complexity and interpretability. Meanwhile, LLMs have the capability to produce natural language explanations, making them more user-friendly and convincing. In this context, we explore the use of an LLM as a surrogate model for explainability in recommender models. We start with a \textbf{behavior alignment} approach, where the LLM is fine-tuned to predict items based on user profiles, closely mirroring the recommendation model's output. While this method provides useful insights, it does not delve into the internal logic of the model. 
To address this, we propose \textbf{intention alignment}, wherein the LLM learns to process the recommender model's embeddings. Similar to how vision-language multimodal models process visual data, this approach aims to enable the LLM to understand the information within user/item embeddings, allowing it to explain the reasoning behind a recommender model's suggestions.
We find that combining these two methods into a \textbf{hybrid alignment} strategy, which incorporates both textual information and embeddings, can more effectively address interpretation inaccuracies and enhance the overall interpretability. This integrated approach combines the benefits of both behavior and intention alignment, providing a stronger and more comprehensive explanation mechanism. 
To implement the three alignment methods --- behavioral, intentional, and hybrid alignment --- we define six tasks to fine-tune an LLM to align with a target recommender model's predictions. These tasks include teaching the LLM to predict the next item a user may like, learning to rank items, classifying interests, detailing item characteristics, maintaining general intelligence through ShareGPT training, and reconstructing user history for intention alignment.  This comprehensive training regimen equips the LLM to replicate the recommender model's logic.  Thus, together with the LLM's own reasoning capabilities and world-knowledge, LLMs can generate model explanations with higher fidelity and robustness in the recommendation scenario. For more technical details and evaluations, please refer to \cite{lei2023recexplainer}.

\section{Evaluator}\label{evaluator}
RecAI provides a tool for automatic evaluation across five key dimensions:

\textsl{Generative recommendation}. LLM-based RSs enable natural language engagement, which can occasionally result in item names being generated with minor inaccuracies, such as incorrect punctuation. To accommodate these potential discrepancies, we employ fuzzy matching to ensure our name validation process remains adaptable without being too strict.

\textsl{Embedding-based recommendation}. RecAI evaluator supports embedding-based matching models like our RecLM-emb or OpenAI's text embedding API\footnote{\url{https://platform.openai.com/docs/guides/embeddings}}. 
    Once user/item embeddings are inferred, the subsequent evaluation procedure aligns with the conventional evaluation process.

\textsl{Conversation}. We assess conversational recommendation efficacy through a GPT-4-powered user simulator that engages with the system to solicit item suggestions. System performance is gauged by its success in referencing the simulator's target items during the interaction.

\textsl{Explanation}. The system delivers explanations for its recommendations, which are then evaluated by an independent LLM like GPT-4, serving as a judge to appraise the informativeness, persuasiveness, and helpfulness of these explanations.

\textsl{Chit-chat}. Users might initiate non-recommendation dialogues, like asking "how to write a research paper." The RS is expected to adeptly manage such inquiries. An LLM, such as GPT-4, critiques the system's replies for their helpfulness, relevance, and thoroughness.

We measure the first three dimensions using NDCG and Recall metrics compared to ground truths. For Explanation and Chit-Chat, we utilize pairwise comparisons for a solid evaluation, where a judge contrasts outputs from two models, tallying wins, losses, and ties to gauge overall performance.

\section{Conclusions}
We present RecAI, a toolkit designed to leverage LLMs to forge recommender systems that emulate human-like interactions. RecAI is structured around multiple pillars, each aimed at addressing a variety of real-world applications through diverse techniques.  For instance, engineers aiming to evolve their industrial recommender systems into conversational interfaces can deploy the Recommender AI Agent framework, thus preserving the value of their existing recommender models. Researchers looking to rapidly develop a conversational recommender system with minimal costs might opt for the Chat-Rec framework, integrating RecLM-emb for retrieval and RecLM-gen as the generative LLM.

We anticipate that the next generation of recommender systems, powered by LLMs, will offer increased versatility, interactivity, and user control. We hope RecAI can accelerate this transformative process, providing the tools necessary for the industry and academia to build more sophisticated, engaging, and responsive recommendation systems.

\bibliographystyle{ACM-Reference-Format}
\bibliography{myref}


\begin{thebibliography}{7}


\ifx \showCODEN    \undefined \def \showCODEN     #1{\unskip}     \fi
\ifx \showDOI      \undefined \def \showDOI       #1{#1}\fi
\ifx \showISBNx    \undefined \def \showISBNx     #1{\unskip}     \fi
\ifx \showISBNxiii \undefined \def \showISBNxiii  #1{\unskip}     \fi
\ifx \showISSN     \undefined \def \showISSN      #1{\unskip}     \fi
\ifx \showLCCN     \undefined \def \showLCCN      #1{\unskip}     \fi
\ifx \shownote     \undefined \def \shownote      #1{#1}          \fi
\ifx \showarticletitle \undefined \def \showarticletitle #1{#1}   \fi
\ifx \showURL      \undefined \def \showURL       {\relax}        \fi
\providecommand\bibfield[2]{#2}
\providecommand\bibinfo[2]{#2}
\providecommand\natexlab[1]{#1}
\providecommand\showeprint[2][]{arXiv:#2}

\bibitem[Gao et~al\mbox{.}(2023)]%
        {gao2023chat}
\bibfield{author}{\bibinfo{person}{Yunfan Gao}, \bibinfo{person}{Tao Sheng}, \bibinfo{person}{Youlin Xiang}, \bibinfo{person}{Yun Xiong}, \bibinfo{person}{Haofen Wang}, {and} \bibinfo{person}{Jiawei Zhang}.} \bibinfo{year}{2023}\natexlab{}.
\newblock \showarticletitle{Chat-rec: Towards interactive and explainable llms-augmented recommender system}.
\newblock \bibinfo{journal}{\emph{arXiv preprint arXiv:2303.14524}} (\bibinfo{year}{2023}).
\newblock


\bibitem[Hou et~al\mbox{.}(2023)]%
        {hou2023large}
\bibfield{author}{\bibinfo{person}{Yupeng Hou}, \bibinfo{person}{Junjie Zhang}, \bibinfo{person}{Zihan Lin}, \bibinfo{person}{Hongyu Lu}, \bibinfo{person}{Ruobing Xie}, \bibinfo{person}{Julian McAuley}, {and} \bibinfo{person}{Wayne~Xin Zhao}.} \bibinfo{year}{2023}\natexlab{}.
\newblock \showarticletitle{Large language models are zero-shot rankers for recommender systems}.
\newblock \bibinfo{journal}{\emph{arXiv preprint arXiv:2305.08845}} (\bibinfo{year}{2023}).
\newblock


\bibitem[Huang et~al\mbox{.}(2023)]%
        {huang2023recommender}
\bibfield{author}{\bibinfo{person}{Xu Huang}, \bibinfo{person}{Jianxun Lian}, \bibinfo{person}{Yuxuan Lei}, \bibinfo{person}{Jing Yao}, \bibinfo{person}{Defu Lian}, {and} \bibinfo{person}{Xing Xie}.} \bibinfo{year}{2023}\natexlab{}.
\newblock \showarticletitle{Recommender ai agent: Integrating large language models for interactive recommendations}.
\newblock \bibinfo{journal}{\emph{arXiv preprint arXiv:2308.16505}} (\bibinfo{year}{2023}).
\newblock


\bibitem[Lei et~al\mbox{.}(2023)]%
        {lei2023recexplainer}
\bibfield{author}{\bibinfo{person}{Yuxuan Lei}, \bibinfo{person}{Jianxun Lian}, \bibinfo{person}{Jing Yao}, \bibinfo{person}{Xu Huang}, \bibinfo{person}{Defu Lian}, {and} \bibinfo{person}{Xing Xie}.} \bibinfo{year}{2023}\natexlab{}.
\newblock \showarticletitle{RecExplainer: Aligning Large Language Models for Recommendation Model Interpretability}.
\newblock \bibinfo{journal}{\emph{arXiv preprint arXiv:2311.10947}} (\bibinfo{year}{2023}).
\newblock


\bibitem[Lei et~al\mbox{.}(2024)]%
        {lei2024aligning}
\bibfield{author}{\bibinfo{person}{Yuxuan Lei}, \bibinfo{person}{Jianxun Lian}, \bibinfo{person}{Jing Yao}, \bibinfo{person}{Mingqi Wu}, \bibinfo{person}{Defu Lian}, {and} \bibinfo{person}{Xing Xie}.} \bibinfo{year}{2024}\natexlab{}.
\newblock \bibinfo{title}{Aligning Language Models for Versatile Text-based Item Retrieval}.
\newblock
\newblock
\showeprint[arxiv]{2402.18899}~[cs.IR]


\bibitem[Lu et~al\mbox{.}(2024)]%
        {lu2024aligning}
\bibfield{author}{\bibinfo{person}{Wensheng Lu}, \bibinfo{person}{Jianxun Lian}, \bibinfo{person}{Wei Zhang}, \bibinfo{person}{Guanghua Li}, \bibinfo{person}{Mingyang Zhou}, \bibinfo{person}{Hao Liao}, {and} \bibinfo{person}{Xing Xie}.} \bibinfo{year}{2024}\natexlab{}.
\newblock \bibinfo{title}{Aligning Large Language Models for Controllable Recommendations}.
\newblock
\newblock
\showeprint[arxiv]{2403.05063}~[cs.IR]


\bibitem[Yao et~al\mbox{.}(2023)]%
        {yao2023knowledge}
\bibfield{author}{\bibinfo{person}{Jing Yao}, \bibinfo{person}{Wei Xu}, \bibinfo{person}{Jianxun Lian}, \bibinfo{person}{Xiting Wang}, \bibinfo{person}{Xiaoyuan Yi}, {and} \bibinfo{person}{Xing Xie}.} \bibinfo{year}{2023}\natexlab{}.
\newblock \showarticletitle{Knowledge Plugins: Enhancing Large Language Models for Domain-Specific Recommendations}.
\newblock \bibinfo{journal}{\emph{arXiv preprint arXiv:2311.10779}} (\bibinfo{year}{2023}).
\newblock


\end{thebibliography}

\end{document}